\documentclass[twocolumn,11pt]{article}
\usepackage{graphicx}
\usepackage{amsfonts}
\usepackage{amssymb}
\usepackage{amsbsy}
\input epsf.sty
\renewcommand{\vec}[1]{\boldsymbol{#1}} 
\def\ll{\label}
\def\re{\ref}

\def\r1{(\ref{$1})}

\def\ba{\begin{array}{c}}

\def\ea{\end{array}}

\def\si{\sigma}

\def\ov{\over}
\def\ha{{1\over 2}}

\def\l{\left}
\def\l({\left(}
\def\r){\right)}
\def\r{\right}

 \def\be{\begin{equation}}
\def\bc{\begin{center}}
\def\ec{\end{center}}
\def\bit{\begin{itemize}}
\def\eit{\end{itemize}}
\def\ee{\end{equation}}
\def\ed{\end{document}}
\def\bea{\begin{eqnarray}}
\def\eea{\end{eqnarray}}
\def\efr{\end{flushright}}

\begin{document}
\noindent {\Large{\bf
 Possibilities of a classical alternative to a quantum computer}}
\vskip 1cm
{\small 
\noindent
{\bf
Anjan Kundu}
 \footnote {email: anjan@tnp.saha.ernet.in} \\  
 \\
{\it  Theory Group.  Saha Institute of Nuclear Physics,  
 1/AF Bidhan Nagar, Calcutta 700 064, India.
 }

%
\vskip 1 cm

\hrule \vskip .6cm
\noindent {\bf
 The dramatic increase in the efficiency of a quantum computer  over a
 classical computer $^{1,2}$, raises a natural question asking, how much of
 this success could be attributed to its quantum nature and how much to its
 probabilistic content. To highlight this issue, we put forward the novel
 idea of a possible chemical computer driven by reaction-diffusion (RD)
 processes$^{3-5}$ based on a probabilistic but classical approach.  Such
 computers, obeying non-equilibrium statistical mechanics, can describe
 superpositions of empty and filled states with certain probabilities.  With
 these {\it probit} states serving as computational basis states, such RD
 computers with operations satisfying a necessary semi-group property
 could  mimic some well known quantum logic gates and carry out
teleportation like procedure using entangled states, believed to be a
 prerogative of the quantum world. Moreover, assuming a nonlinear extension
the RD computers could be used for cloning of arbitrary states, which is a
famous forbidden operation in standard quantum computation.}
 


  It is  important  to compare the performance of a quantum computer 
 (QC) with  that of a classical but stochastic computer,
 to understand more clearly the reason behind  the
success of a QC.
Though the randomised algorithm of Solovay-Strassen has prompted 
 generalisation of the universal Touring machine to a probabilistic one
 $^{1,2}$, apparently  not much attention  
  has  been paid  in suggesting  
 a    computer, driven by 
probabilistic processes, as a possible  alternative to a QC.
  We observe however that, such classical stochastic events, could be
realised in
 physical systems controlled 
 by   RD processes$^{3-5}$, 
 which can describe a { state} as the { superposition} $|P>=p_0
|0> +p_1 |1>,$ with $ \ p_0+p_1=1 $, where $ p_0$ is the probability with
which  an {empty} state $ |0> $  ($\circ$)
 can occur, while
$ p_1$ is the corresponding probability of $ |1> $, i.e of the state 
occupied by a single particle ($\bullet $). 
Such  processes may be involved  in an one dimensional 
array of $N$ lattice sites and along with
  diffusion to the right  or to the left ($ \bullet + \circ 
\leftrightarrow
\circ + \bullet $), can describe
reactions like
  coagulation, decoagulation
($ \bullet + \bullet 
\leftrightarrow \bullet $) , birth and death ($  \circ 
\leftrightarrow
 \bullet + \bullet $),
 occurring with certain probabilities.
  The evolution of these probabilities is 
 governed by a  master equation
$\partial_t p_{\vec{ \sigma}}=H_{\vec {\si}\vec{ \tau}}
p_{\vec {\tau}}$, 
where $p_{\vec{\sigma}}$
 is the probability of the system to be 
in  state $|{\vec{\sigma}}=(\sigma_1,\sigma_2 \ldots, \sigma_N)>$,
 defined by a configuration stretched over $N$ lattice sites with
  $\sigma_i$'s being
 $0$ or $1$. The  dynamical operator $H_{\vec{\si}\vec{ \tau}} $ 
expressed through  transition rates between the configurations 
${\vec{\sigma} }$ and ${\vec{ \tau }}$
 is assumed to be a linear operator determined by the physical
processes
 involved.  By  introducing $|P>=\sum_{\vec{\sigma}} p_{\vec{\si}}
 |\vec{\si}> $, 
the evolution equation
   can be cast as a real Schr\"odinger equation   
$\partial_t |P>= H|P> $, with
the  state evolving in a 
finite time  as  $|P(t)>=U(t-t_0)|P(t_0)>
$.
The linear evolution  operator 
$U$ with  only  positive and real matrix elements has an important
 constraint 
\be\sum_i^NU_{ij}=1, \ j=1, \ldots ,N ,\ll{Urd}\ee
imposed by
 the conservation of total probability:  
$\sum_{{\bf\sigma}}p_{{\bf\sigma}}=1$.
 We  call any matrix satisfying condition 
(\re{Urd}) as the RD matrix.

We can notice   the striking  similarity between  
  these  relations  and those of the quantum mechanics, where only the
 superposition coefficients are the   probability amplitudes,
the evolution equation
is the complex Schr\"odinger equation
and the evolution operator is  a  unitary matrix.
 This similarity,  though formal,  
 should be motivation  enough to explore the idea of a    RD computer  
 for a possible alternative to a QC.

Note that,  for an  operation to  qualify  for any sequential
computation, a primary requirement is that, it must exhibit at least a
semi-group property.
The quantum  computation is  guaranteed to have the semi-group 
property, following from its  
 unitary group transformation linked  to the 
 reversibility of quantum dynamics.
Returning to the   RD processes,
  we find  rather unexpectedly that,   
the defining  constraint (\re{Urd})    is    preserved
  under evolution: 
$\sum_iU_{ij}(t-t_0)=\sum_{k}(\sum_{i}
(U_{ik}(t-t_1))U_{kj}(t_1-t_0)=\sum_{k}
U_{kj}(t_1-t_0)=1$
 and since $U=I $   satisfies  (\re{Urd}), the set of RD matrices defining
  RD operations forms a semi-group. Moreover, since the inverse
   operation $(U)^{-1}$,
if it exists,  should also satisfy  (\re{Urd}), some of the RD operations
  might form a group. However, in general the required positivity condition
on
 the matrix elements are not  preserved under inverse operation and the
 group structure is obtained formally, only if this positivity condition is
 relaxed. We find that, such a group with $N \times N$ matrices with real
 entries satisfying (\re{Urd})
would involve $N(N-1)$ generators and  act as a transformation group 
on a plane   embedded  in a $N$-dimensional Euclidean space,
 with defining relations
$\sum_i^Nx_i=1$ on its coordinates. Such groups are
simplest examples of   algebraic groups.
Interestingly, the matrices satisfying additional constraints similar to
(\re{Urd}) on their rows
:$ \sum_j^NU_{ij}=1, \ i=1, \ldots ,N  $ constitute a subgroup of
dimension $(N-1)^2$, which acts as a stationary  group  keeping the 
point  with
$x_1=x_2= \ldots =x_N={1 \ov N}$  on the plane fixed under its action.

Therefore,  we may apply in principle the RD processes for computation
 purposes
and in analogy with a qubit  call the elementary information in 
 a  RD  computer, based on   probabilistic approaches, as  {\it probit}. 
In contrast to the qubits belonging to a Hilbert space 
with the state vectors describing a unit sphere,
the  probit states would
  belong to a  real   vector space, with the tip of the state vectors moving
 on a plane limited between positive coordinate axes of unit lengths.
The action of   RD matrices   
 keeps this plane invariant by transforming a point to another point on the
 same plane.  For single probit states
the RD operators are $2 \times 2$ matrices defined by two independent
parameters  $U_{00}=a, U_{01}=b $. We find a
 remarkable property that 
$\lim_{n \to \infty} (U(a,b))^n = U(a_\infty,a_\infty)$, where 
$a_\infty = {b \ov 1-a+b}$.
 That is   on repeated   operation by an 
 arbitrary  RD  matrix,  all state vectors   end up in an
equilibrium
state $|s>=a_\infty |0> +(1-a_\infty) |1>$ 
, irrespective  of the probit state we 
start with (see Fig 1).

   \rotatebox{-90}{\epsfxsize=150pt \epsfbox{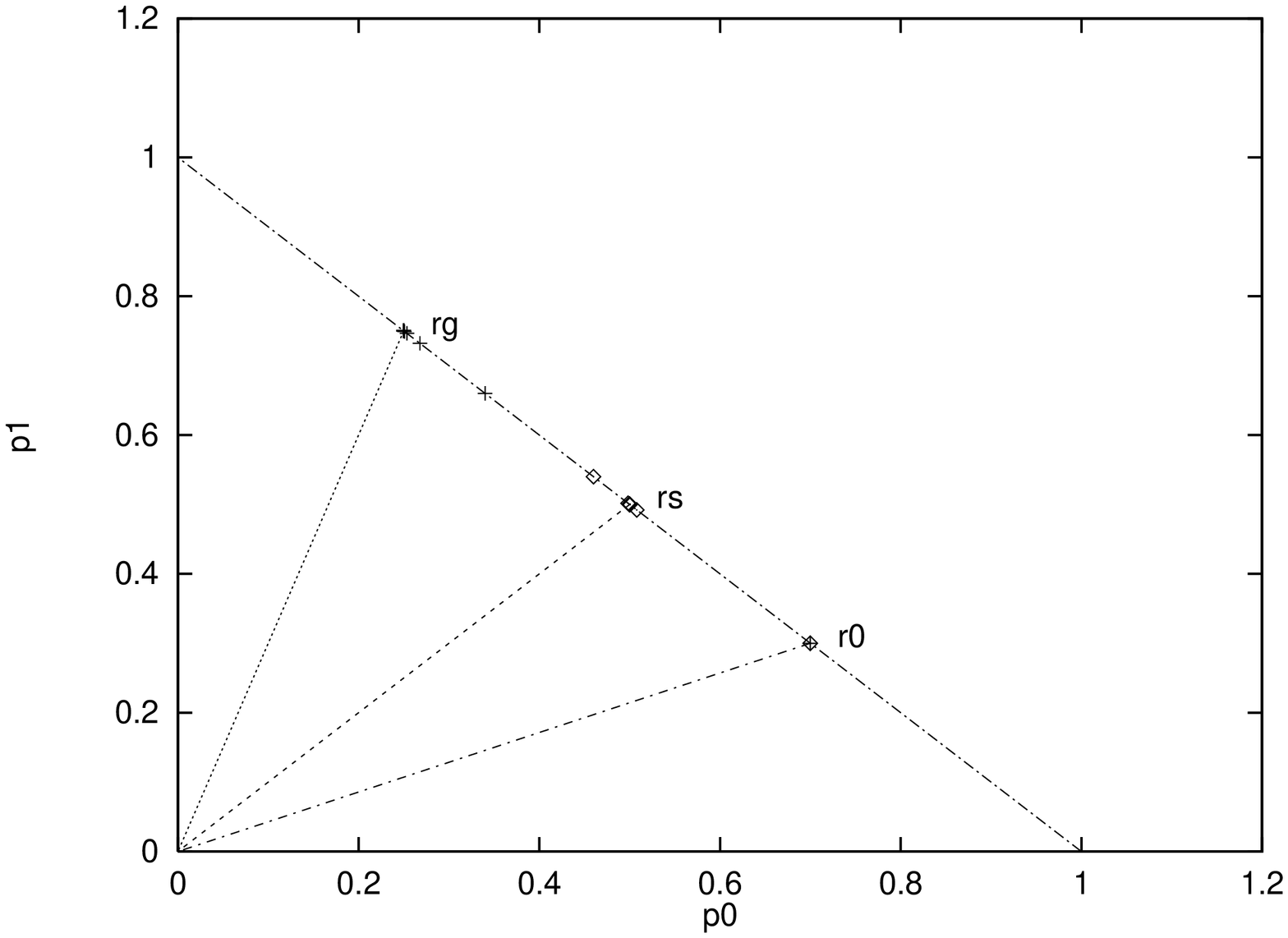}} 

\vskip .5cm

\noindent {\bf Figure 1} {\tiny  Single probit  states 
under action of RD operations,  confirming  also the semi-group property
of RD matrices.  Under repeated operations by $U(a,b)$
 all vectors  from arbitrary   initial position $r_0$
 evolve into a fixed point $r_g=(a_\infty,1-a_\infty)$ (see text).
For  symmetric matrices, this  
fixed point $r_s=(\ha , \ha)$ becomes independent of even  the operating RD
matrix.  
In the figure  particular values  $a=0.4, b=0.2$,
  $r_0=(0.7,0.3), r_g=(0.25,0.75)$, and for the  
 symmetric case   $a=0.4,\  b=0.6$ have been considered.
 }

For   demonstrating  the action of such computers we  start  
with simplest RD  operations, which could act like  { logic gates}
 and transform
computational probit states $|0>, |1>$
 to mixed states: $|0> \to U_{00} |0> + U_{10}  |1>, \ 
|1> \to  U_{11} |1> +  U_{01} |0> $,
 with $U$ being a RD matrix satisfying (\re{Urd}). 
Notice that, if $ U_{01}=  U_{10}=1$,
  an  empty state  ends up giving birth to  a particle,
 while a    filled state  becomes empty through a death process,
 i.e. the   operation acts  like
a  NOT-gate,
 similar to a quantum logic (QL) $X$-gate.
 Therefore, an arbitrary single probit  state $|P>=p_0
|0> +p_1 |1>$
 after passing through such a NOT-gate,  due to the
linearity of  RD operations would yield 
   $ |\tilde P>=p_1
|0> +p_0 |1>$, i.e the probabilities of computational states 
would toggle. 
Note  that, other popular QL gates like $Z$-gate and the
 Hadamard-gate having
negative matrix elements$^1$ 
are not allowed  here 
 and  this  particular limitation,  as we see below,  
prevents  a RD computer to match a QC in solving some important
 algorithmic problems. 
Reciprocally however,  while
only  the NOT-gate and the identity operation 
can have  equivalent QL gates, 
 all other values of $U_{ij}$ 
 would represent
 new types of single probit gates achievable  by RD computers alone.
For example, the matrix with  all $U_{ij}=\ha $, having no inverse,
 transforms both
$|0>, |1>$ to the same uniform state $|u>=\ha (|0>+|1>)$.
 We  use below this uniforming 
 RD gate  
in designing teleportation of probit states. 
 
Focusing now on two-probit operations
we show that,  { C-NOT} and { Swap} gates
  can also be defined   through RD computation similar to QC.  
 Defining in analogy with QL gates$^1$
   the first probit as the control state 
 $|x>$ and the
 second  one  as
the target state   $|y>$, we may construct    
 a C-NOT gate   through    RD operations, such that,
 if $|x=0>$ is an  empty state, 
 nothing happens to  $|y>$,
 but if a particle is present 
 in $|x=1>$, the computational state  $|y>$ is negated
 through the NOT gate presented above.
We  check easily that the matrix  
$U^{CNOT}$ representing this operation 
 satisfies the RD condition (\re{Urd}). Observe also that, 
  the
 control  states can be copied by fixing 
$y=0$, only if they are in the 
computational basis and not in the arbitrary state $|P>$. This
 indicates  that
the no-cloning theorem, a  well known no-go theorem in QC, is
 valid also for the RD
computers. However we will see below that, allowing nonlinear RD processes 
one could    bypass this theorem. 

We may  construct the swap logic gate P
driven by a RD computer, again in  analogy with the  QL gate.
 The action of such a  gate through  RD operations would
mean, when the neighbouring two sites of the system are the same,
 i.e, either both empty or both filled--
nothing changes. However,
 when a particle appears only at one site, it
diffuses to the empty site, either to the right or to the left.
We  check again that this 
is a permitted 
RD matrix operation satisfying (\re{Urd}).

More interesting f-gates$^1$  can be constructed as well in a RD computer,
where a function $x \to f(x) $ mapping the values $0,1$ to themselves, 
  is induced by   a logic gate,  that changes  the target state $|y> \to
|y+f(x)>$, depending on the nature of the control state $|x>$.
We see immediately that similar to the C-NOT gate
 all such f-gates satisfy the RD condition (\re{Urd}).
Note that, for all  above logic gates like NOT, C-NOT, Swap and f-gates
 the related  RD matrices have well-defined inverses
 and therefore each of these RD operations
 forms  a group. Moreover such RD operations formally coincide with the
corresponding
 QL gates exhibiting unitarity and  therefore the  RD computers 
using these gates can serve as possible alternative to QC for 
related computations.

Another operation where RD computation  can mimic a QC is the
celebrated teleportation problem$^1$. Here also the
probabilistic logic can go parallel to the quantum logic and 
 as in the quantum
case we may consider that, Alice and Bob share an entangled two-probit
Bell-like  state
$ \ha(|00>+|11>)$.
 For teleporting a RD state $|P>=p_0|0>+p_1|1>$, Alice
passes it as a control state through a C-NOT gate together with the first
probits of the shared entangled state in her possession as the target,
 similar to the procedure adopted in  QC. This gives the three
probit state
  $\ha(p_0(|000>+|011>)+p_1(|110>+|101>))$, the first probits of 
 which  are then sent by 
  Alice   through a uniforming RD gate, elaborated above, 
yielding the final state
${1 \ov 4}\{(|00>+|10>)|P>+(|11>+|01>)|\tilde P>\}$
with $|\tilde P>$ being the toggled state.
Therefore, if now Alice in her measurement of her two probit states
 finds either $|00>$ or $|10>$
the state $|P>$ itself has been teleported, while if the result is
either  $|11>$ or $|01>$,  the  toggled-state $|\tilde P>$
has been sent, in which case  the original state
is recovered  by passing  it     through a
NOT-gate. The whole process could also be repeated using 
another Bell-like state $ \ha(|01>+|10>)$, which is the only other
 entangled state possible  in RD computation. 
 Thus  it appears that,  quantum teleportation like scenario
 could be  reproduced 
within the classical RD processes.

Limitations of RD computers would show up 
in dealing certain algorithmic problems, which renders 
efficient solutions in QC.
 Deutsch problem is to tell
 whether a Boolean function $f$
  is  constant (i.e. $f(0) =  f(1)$ ) or  balanced ( i.e.
$f(0) \neq f(1)$) by testing   the outcome of an operation, while its
  Jozsa generalisation    
 asks  the same question (with slightly different definitions) for  
 a N-bit state$^1$.
For testing these problems in a RD computer we may take 
the uniform state $|u>$ as the control and $|0>$ as the target
 and  see that,  this pure input 
state   after passing through the f-gate
remains pure (either $|u>|0>$ or $|u>|1>$) if the gate is constant,
 but
  turns into an entangled state  in the balanced case. 
Therefore if somehow we can  test for the  entanglement,
 that would solve our problem. 
However, how to do this efficiently  is not clear at this stage. 
Similarly,  Grover's search algorithm$^{1,2}$ in RD computation may
 also start with  the 
uniform state $|u>$ 
   passing  through the oracle (initialised in the state $|0>$).
If the oracle is 
attached to a  search function $f$   yielding $f(X)=1$ for the 
solution set $X$ and $0$, otherwise, 
one  naturally gets 
 at the output a superposition of  orthogonal states $ \sum_X|X>|1>+
\sum_{X'}|X'>|0>$.
 Therefore,  though  the solution  states $|X>$ are
 distinguishable  theoretically  from the other states $|X'>$,
 apparently  in no way we can      
   detect them  to match the
efficiency of a QC. In case of Shor's factorisation algorithm$^1$, the case is
even worse, since  from the very beginning
 it  uses complex  superposition
coefficients, which are not permitted in RD operations.

In spite of these deficiencies of the proposed RD computer,
 its  domain permissible beyond   
 reversible processes might give
 some generalities, as we have seen with the uniforming
RD gate. 
 Another possibility  of extending  RD processes with
important potential applications  is to    induce 
 nonlinear operations.
Let us suppose that we intend to clone an arbitrary state $|P>=p_0
|0> +p_1 |1>.$   Since we have the
indication that this is not possible with   linear RD processes, 
  we   allow   the RD operator $U=I \otimes U(P) $ to depend on 
the initial state $|P>$. If it acts  on the  product state
$\ha|P>(|0>+|1>)$, simple calculations  reveal that,
one could obtain   the cloned state $|P>|P>$ at the output point,
 by requiring the nonlinear
 dependence as $ U(P)_{00}+ U(P)_{01}=2p_0$.  and $
U(P)_{10}+ U(P)_{11}=2p_1$. We notice that the RD matrix condition
(\re{Urd})  still holds due to $p_0+p_1=1$.
 Note that, since in RD processes the
 transition rates  are determined mostly
phenomenologically, the assumption of nonlinearity 
should not lead to
any serious consequences, like violation of
causality, faster than light propagation etc. as one must expect 
with similar   nonlinear assumption
in 
quantum computations$^1$. 

Therefore  similar to quantum computers, our proposed classical 
computers based on 
reaction-diffusion processes and using  probabilistic probit states 
can   construct many important
 logic gates and  carry out teleportation 
like processes, while through their nonlinear extension  
   the cloning of mixed
states could be made possible.
  Though the present work
is purely theoretical,  this novel
 concept of RD computation, being classical in nature, might be easier than
a quantum computer to realise and handle in practice.
\vskip .5cm

\hrule
\vskip .2 cm
{1.} Nielsen, M. A., nd Chuang, I. L.,
 { \it Quantum Computation and Quantum Information}
( Cambridge Univ Press, Cambridge, 2000)

2.  Leuenberger, M. N., and  Loss, D.,
 { Quantum computing in molecular magnets}.
{\it Nature},{\bf 410}, 789-793 (2001)

3. Glauber, R.J., Time dependent statistics of the Ising model.
 {\it J. Math. Phys.}, {\bf 4}, 294-307, (1963) 

4. {\it Nonequilibrium Statistical Mechanics  in One Dimension} (
Privman V. (ed.), Cambridge Univ Press, Cambridge 1996)


{5.} Henkel, M.,   Reaction-diffusion processes and their connection with
integrable quantum spin chains. In {\it Classical and Quantum Nonlinear
Integrable Systems}(ed. Kundu A.), 256-287 , (IOP publ., Bristol, 2003) 
\vskip .5cm

\noindent {\bf Acknowledgments}
I sincerely thank Prof A Chatterjee for explaining  various nuances in 
quantum computation and 
  carefully reading  the manuscript and  Dr Pijush
Ghosh for an useful technical suggestion.
} \end{document}